# Anomalous Hall effect induced spin Hall magnetoresistance in an antiferromagnetic $Cr_2O_3$/Ta bilayer

Yang Ji,[1] J. Miao,[1,a)] K. K. Meng,[1] X. G. Xu,[1] J. K. Chen,[1] Y. Wu,[1] and Y. Jiang[1,b)]

*Beijing Advanced Innovation Center for Materials Genome Engineering, School of Materials Science and Engineering, University of Science and Technology Beijing, Beijing 100083, China*

**ABSTRACT**

The spin Hall magnetoresistance (SMR) and anomalous Hall effect (AHE) are observed in a $Cr_2O_3$/Ta structure. The structural and surface morphology of $Cr_2O_3$/Ta bilayers have been investigated. Temperature dependence of longitudinal and transverse resistances measurements confirm the relationship between SMR and AHE signals in $Cr_2O_3$/Ta structure. By means of temperature dependent magnetoresistance measurements, the physical origin of SMR in the $Cr_2O_3$/Ta structure is revealed, and the contribution to the SMR from the spin current generated by AHE has been proved. The so-called boundary magnetization due to the bulk antiferromagnetic order in $Cr_2O_3$ film may be responsible for the relationship of SMR and AHE in the $Cr_2O_3$/Ta bilayer.

a) Electronic mail: j.miao@ustb.edu.cn
b) Electronic mail: yjiang@ustb.edu.cn

It is known that the spin-orbit coupling (SOC) provides a mechanism to couple charge and spin of electrons and results in spin Hall effect (SHE) in non-magnetic (NM) heavy metals.[1-4] Owing to the spin-orbit interaction, the charge current can be converted into spin current in perpendicular direction in NM layer. As a SOC-related phenomenon, the SHE describes the changing of charge current and pure spin current without ferromagnets (FM).[5,6] On the other hand, the inverse SHE (ISHE) indicated that the pure spin current can be converted back into charge current in perpendicular direction. Several effects, such as spin pumping[7,8], spin transfer torque[9], and spin Seebeck effect[10,11], are correlated with the SHE and the ISHE.

Among those, the Spin Hall magnetoresistance (SMR) is a special magnetoresistance that no charge current flows in the FM layer, which depends on the relative orientations between the magnetic moments of FM layer and the spin polarization direction of NM. SMR has been observed in ferromagnetic insulators,[12-16] for example, H. Nakayama *et al*. got 0.01% MR ratio in $Y_3Fe_5O_{12}$/Pt bilayer and firstly defined it as SMR.[12]

Recently, SMR were predicted to be observed in antiferromagnetic insulators.[17,18] Han *et al*. reported that SMR signal was existed in an antiferromagnetic insulator $SrMnO_3$.[19] In addition, SMR has also been reported in antiferromagnetic $Cr_2O_3$[20] and NiO.[21,22] Among these, $Cr_2O_3$ is an uniaxial antiferromagnetic material with Néel temperature $T_N \approx 308$ K. (0001)-oriented $Cr_2O_3$ has an layer of net surface magnetic moment whereas the bulk keeps antiferromagnetic.[23-25] As a result, the interfacial effect, e.g. SMR, can explore the magnetic properties due to the magnetic textures and surfaces of $Cr_2O_3$.

In this work, both SMR and AHE signals are observed in $Cr_2O_3$ (15)/Ta (5) bilayer. To exclude the magnetic proximity effect (MPE) of platinum, tantalum was adopted as spin source. Our result is different from that previous work of $Cr_2O_3$/W structure,[20] in which $Cr_2O_3$ is only treated as an antiferromagnet. The so-called boundary magnetization in $Cr_2O_3$ layer is essential to obtain the AHE, which is due to the bulk antiferromagnetic order in $Cr_2O_3$ film.[23] As a result, the existence of AHE results in spin accumulation at the interface, which contributes to the SMR together with spin current by SHE. Recently, Yang *et al.*[26] observed a magnetoresistance induced by AHE in ferromagnetic metals without NM and defined it as anomalous Hall magnetoresistance (AHMR), which has the same angular dependence as SMR. The difference between SMR and AHMR is the source of the spin current, i.e., the former is originated from SHE in NM; the latter is originated from AHE in FM. Here, because the contributions from SHE and AHE exist in our structure, both of them are termed as SMR in this work. However, SMR induced by AHE has not investigated in the magnetic insulators yet.

In our experiments, 15 nm thickness $Cr_2O_3$ thin films were grown on $Al_2O_3$ (0001) substrates by using pulse laser deposition (PLD) in oxygen ambience. The PLD chamber utilized a KrF laser of 248 nm wavelength. The laser energy density was kept at 1.8 J/cm$^2$ with a repetition rate of 3 Hz and the target to substrate distance was maintained at 5 cm during the deposition. The base pressure of the chamber before deposition was less than $5\times10^{-5}$ Pa. The thin film deposition was performed at 500 ℃ and 5 Pa oxygen partial pressure. The cooling process was carried at 10 ℃/min to room temperature under $10^4$ oxygen partial pressure. After deposition, a 5 nm Ta was in-situ grown on the $Cr_2O_3$ by magnetron sputtering, where the base pressure of the chamber was less than $8\times10^{-6}$ Pa. Finally, a Hall bar for electrical measurements was prepared by electron beam lithography and Ar ion etching,

The geometry and SEM image for Hall-bar magnetotransport measurements are shown in Fig. 1(a) and 1(b). The size of Hall-bar is 400 × 40 μm$^2$ and a constant channel current **I** of 20 μA flows along the **x** direction. The phase structure of the film was determined by X-ray diffraction ω-2θ scans using M21XVHF22 X-ray diffractometer with Cu/Ka. In addition, the surface morphology was checked by using a Scanning Probe Microscopy (Bruker Icon). The magnetotransport measurements were measured using a Versalab Cryogenic system.

Fig. 1(c) shows the ω-2θ (out-of-plane) XRD scan for the Al$_2$O$_3$ (0001)/Cr$_2$O$_3$ (15 nm) sample. As shown, the (0006) Cr$_2$O$_3$ and (00$\underline{12}$) Cr$_2$O$_3$ peaks are observed at 39.8° and 86.1° respectively. It has been reported that (0006)-oriented Cr$_2$O$_3$ film is α-Cr$_2$O$_3$, which is antiferromagnetic phase.[23,25] No other phases (for example, CrO$_2$ or CrO$_3$) or alternative Cr$_2$O$_3$ orientations are observed in the curve, indicated the good epitaxial properties for Cr$_2$O$_3$ film.

Fig. 1(d) shows AFM image of the Cr$_2$O$_3$ film and the roughness of the film ($R_{rms}$) is 0.189 nm. Since the SMR is a type of interface effect, the flatness of Cr$_2$O$_3$ would be critical for the SMR effect.[27,28] Interestingly, the surface of Cr$_2$O$_3$ layer is relatively smooth without any cracks or pinholes, which is beneficial to growing the upper Ta layer.

The schematics of the angular α, β, and γ dependence are displayed in Fig. 2(a). The longitudinal resistance $R_{xx}$ of the Cr$_2$O$_3$ (15)/Ta (5) structure was carried out as a function of magnetic fields **H** along coordinate axis **x**, **y** and **z** at room temperature in Fig. 2(b). The $R_{xx}$ exhibits a typical SMR-like behavior with $R^x \approx R^z > R^y$, which is similar with the behaviors in FM/NM bilayers.[29] As previous reported in Cr$_2$O$_3$/W structure,[20] the magnetic moment of structure cannot be entirely reversed by external magnetic field, which is caused by the antiferromagnetic property of $Cr_2O_3$. The angle-dependent magnetoresistance (ADMR) measurements were carried out in Fig. 2(c), i.e. the $R_{xx}$ dependent on α, β, and γ. Clearly, $R_{xx}$ keeps almost constant with varying γ, confirming that no AMR exists in our structure. Oppositely, with rotating β, the variation of $R_{xx}$ coincides with $cos^2β$, consisting with the SMR behavior. Fig. 2(d) shows ADMR with β rotation at different temperatures, which is similar as that of $Y_3Fe_5O_{12}$/Pt.[12,30] As the temperature decreasing, the SMR ratio of the $Cr_2O_3$/Ta bilayer increases rapidly at first, and drops as lower than 100 K. The physical original is descried as below.

It is known that the spin-dependent parameters are consistent with the thickness of NM. Fig. 2(e) shows the thickness dependence of Ta layer on the SMR ratios of $Cr_2O_3$ layer. Accordingly, the 4 nm Ta is the maximize spin absorption and reflection, which is close to other reports.[31] It is known that the SMR resistivity change can be described as[12,13]

$$\rho_{xx} = \rho_0 - \Delta\rho_1 m_y^2, \quad (1)$$

$$\rho_{xy} = \Delta\rho_1 m_x m_y + \Delta\rho_2 m_z. \quad (2)$$

where the resistivity $\rho_{xx}$ is measured along the direction of the electric-current flow **I** (along the **x** direction); $\rho_{xy}$ is the Hall resistivity component recorded in the plane perpendicular to **I** (along the **y** direction); $\rho_0$ is a constant resistivity offset; $m_i$ (i = x, y, z) is the i-axis component of the moments; $\Delta\rho_1$ (caused mainly by the real part of spin mixing conductance $G_r$)[13] contributes to the conductance modulation that depends on the in-plane component of the magnetization; while $\Delta\rho_2$ (caused mainly by the imaginary part of spin mixing conductance $G_i$, which can be interpreted as an effective exchange field acting on the spin accumulation)[13] contributes only when there is a magnetization component normal to the plane (AHE), as discussed below. In Figs. 3(a)-(c), the evolutions of the $\Delta\rho_1/\rho_0$ is presented as a function of H, applied at different angles α, β, γ. When H rotates with different angle γ, the $\Delta\rho_1/\rho_0$ curve is almost unchanged. Thus the AMR effect in $Cr_2O_3$/Ta bilayer can be excluded again. Figs. 3(a) and 3(b) show the in-plane α and out-of-plane β rotations data of the SMR, which is consistent with SMR theory.[12,13] The longitudinal and transverse resistance ratio can be descried as

$$\frac{\Delta\rho_1}{\rho_0} = \theta_{SH}^2 \frac{\lambda}{d_N} \frac{2\lambda G_r tanh^2\frac{d_N}{2\lambda}}{\sigma+2\lambda G_r coth\frac{d_N}{\lambda}}, \qquad (3)$$

$$\frac{\Delta\rho_2}{\rho_0} = \theta_{SH}^2 \frac{\lambda}{d_N} \frac{2\lambda \sigma G_i tanh^2\frac{d_N}{2\lambda}}{\left(\sigma+2\lambda G_r coth\frac{d_N}{\lambda}\right)^2}. \qquad (4)$$

For a 5-nm-thick β-phase Ta film with the high resistivity approximate 320 μΩ·cm,[3] the experimental for the spin Hall angle $\theta_{SH}$ = 0.03, the spin-flip diffusion length of γ = 1.8 nm,[30] and we get the spin mixing conductance $G_r$ ≈ 1.77 × $10^{14}$ $Ω^{-1}$ $m^{-2}$, which is similar to the YIG/Pt bilayer structure.[12,27,28] Using the experiment result $\Delta\rho_2/\rho_0$ = 3.17 × $10^{-5}$, $G_i$ ≈1.01 × $10^{14}$ $Ω^{-1}$ $m^{-2}$ is extracted from Eq. (4). This gives $G_i/G_r$ = 0.57, which exceeds the ratio 0.05 in YIG/Pt bilayer.[13,32,33] The higher ratio confirms that AHE is not only a derivative of SMR, i.e., spin Hall AHE (SHAHE),[13] but also has a special contribution to SMR.

Fig. 4(a) shows the temperature dependence on the Hall resistance measurements of $Cr_2O_3$ (15)/Ta (5), and the S-like curves indicate the presence of AHE.[34] In general, AHE is an effect existing in ferromagnetic materials, while $Cr_2O_3$ is an antiferromagnetic insulator. To investigate the origin of AHE in $Cr_2O_3$, the magnetization structure of $Cr_2O_3$ should be taken into account. It has been reported that surface magnetization exists at the surface of (0001)-oriented $Cr_2O_3$, namely, the whole film can be divided into surface ferromagnetism region and bulk antiferromagnetism region.[23-25] Due to both SMR and AHE are interfacial effects, the surface magnetization would be dominant at its interface. As a result of relative weak exchange interaction between spins and antiferromagnetic order, its direct interactions on SMR and AHE can be neglected. To confirm the observations, Fig. 4(b) shows the temperature dependence of $\Delta R_{xy}$ ($R_{xy}^{-3T}$ - $R_{xy}^{3T}$) and SMR ratio. It is unambiguous that SMR coincides with $\Delta R_{xy}$ dependent on the temperature, confirming that the spin accumulation plays a dominating role in spin transport at its interface, as similar physical mechanism reported by Yang.[26]

Fig. 5(a) –5(c) show $R_{xy}$ variation in three planes, *xy*, *yz*, and *xz* plane as a function of the angles of α, β, and γ, respectively. A magnetic field of 3 T is applied for all rotations, which is larger than the saturated field ~ 2 T for $Cr_2O_3$ (15)/Ta (5) sample in Fig. 4(a). The variations of $R_{xy}$ dependence on β and γ are due to the $m_z$, whereas the variation of $R_{xy}$ with α is attributed to $m_x m_y$, namely the planar Hall effect (PHE). Noted that the PHE can be attributed to the transverse SMR, which is comparable to that of the AHE. This can explain the large PHE which has been observed in W/CoFeB/MgO.[35] However, $\Delta R_{xy}$ in *yz*, and *xz* plane are twice as much as that in *xy* plane, indicating the easy axis is perpendicular to the film. On the other hand, the second term on the right in Eq. (2) control the magnitude of $\rho_{xy}$ mainly, owing to that $m_z$ is larger than $m_x$ and $m_y$ in α-$Cr_2O_3$. The results are consistent with the magnetic feature of (0001)-oriented $Cr_2O_3$,[23] in which the moments are along the c-axis (0001). Thus, AHE is proved again that it should be a source of spin current and control the magnitude of SMR. Nevertheless, since AHE and SMR have the same trend with temperature in Fig. 4(b), AHE is considered to be a primary role for supplying spin current.

To confirm the angular dependence of $R_{xy}$, the measurement rotating the samples along with the angle α, β, γ under 3 T are shown in Figs. 6(a) - (c), respectively. As shown in Fig. 6(a), when the magnetic field is applied in-plane, the variation trend of $R_{xy}$ is similar to that of YIG/Pt structure,[12] proving that the ferromagnetic component of $Cr_2O_3$ plays a role in ferromagnetic layer. Moreover, the β and γ dependence of $R_{xy}$ are consistent with the Figs. 5(b) and 5(c), respectively. $\Delta R_{xy}$ with β and γ rotations is ~ 0.24 Ω and nearly twice larger than that of α rotation.

In conclusion, a 0.017 % SMR ratio under 3 T at 300 K in a $Cr_2O_3$/Ta bilayer is observed, which is increased to be almost threefold at 100 K. By simultaneous detections on SMR and AHE, the surface magnetization and out-of-plane orientation magnetization of $Cr_2O_3$ is responsible for AHE in this structure. The trends of temperature dependence of both SMR and AHE in $Cr_2O_3$/Ta bilayers are similar. The perpendicular magnetic structure of $Cr_2O_3$ makes AHE become a major source of SMR. Our results indicate that the SMR signal should be originated from both the spin current by SHE in NM and the spin accumulation at interfaces by AHE. Our work can be benefited for the understanding the origin of SMR.

**ACKNOWLEDGEMENTS**

This work was partially supported by the National Basic Research Program of China (No. 2015CB921502), National Science Foundation of China (Nos. 51731003, 11574027, 51671019, 51471029), National Key R&D Program of China (No. 2018YFB0704100), Beijing Municipal Science and Technology Program (Z161100002116013) and Beijing Municipal Innovation Environment and Platform Construction Project (Z161100005016095). J.C., K.M., and Y.W. acknowledge the National Science Foundation of China (Nos. 61674013, 51602022, 61404125, 51501007).

**FIGURE LEGENDS**

*Fig. 1.* (a) A schematic illustration of the Ta Hall bar/$Cr_2O_3$ film sample used for measuring the SMR. (b) Optical image of the Hall bar, in which the length is 400 μm, and the width is 40 μm. (c) XRD 2θ scan from a 15 nm $Cr_2O_3$ film grown on $Al_2O_3$ (0001) substrate. (d) AFM of 15 nm $Cr_2O_3$. The roughness is 0.189 nm.

*Fig. 2.* (a) Notations of different rotations of the angular α, β, and γ. (b) External magnetic field dependence of resistance curve for $Cr_2O_3$ (15)/Ta (5) at 300 K. (c) α, β, and γ dependence of resistance curve for $Cr_2O_3$ (15)/Ta (5) with 3 T at 300 K. (d) β dependence of resistance curve for $Cr_2O_3$ (15)/Ta (5) with 3 T at different temperatures. (e) Longitudinal resistance ratio for $Cr_2O_3$ (15)/Ta (t) with 3 T, in which the Ta thickness t = 3, 4, 5, 7 nm.

*Fig. 3.* (a) Δρ/ρ$_0$ in $Cr_2O_3$ (15)/Ta (5) film as a function of externally applied magnetic field with the different orientation α, β, and γ (see Fig. 2 (a)). The data were taken at T = 300 K.

*Fig. 4.* (a) Hall curves of $Cr_2O_3$ (15)/Ta (5) with the H perpendicular to the film at different temperatures. (b) Temperature dependence of Δ$R_{xy}$ and SMR ratio, in which $\Delta R_{xy} = R_{xy}^{-3T} - R_{xy}^{3T}$.

*Fig. 5.* (a) $R_{xy}$ in $Cr_2O_3$/Ta film as a function of the orientation α, β, and γ of the externally applied magnetic field H (see Fig. 2 (a)). The data were taken at T = 300 K and H = 3 T.

*Fig. 6.* (a) $R_{xy}$ in $Cr_2O_3$ (15)/Ta (5) film as a function of externally applied magnetic field with the different orientation α, β, and γ (see Fig. 2 (a)). The data were taken at T = 300 K.

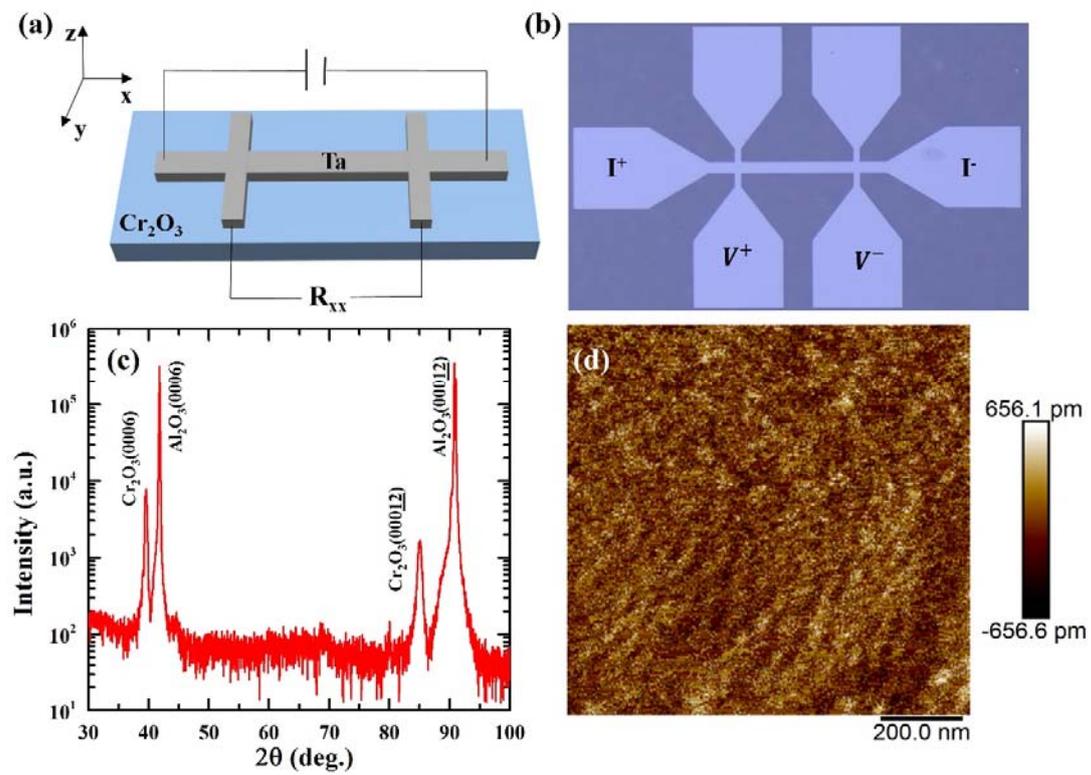

*Figure 1*

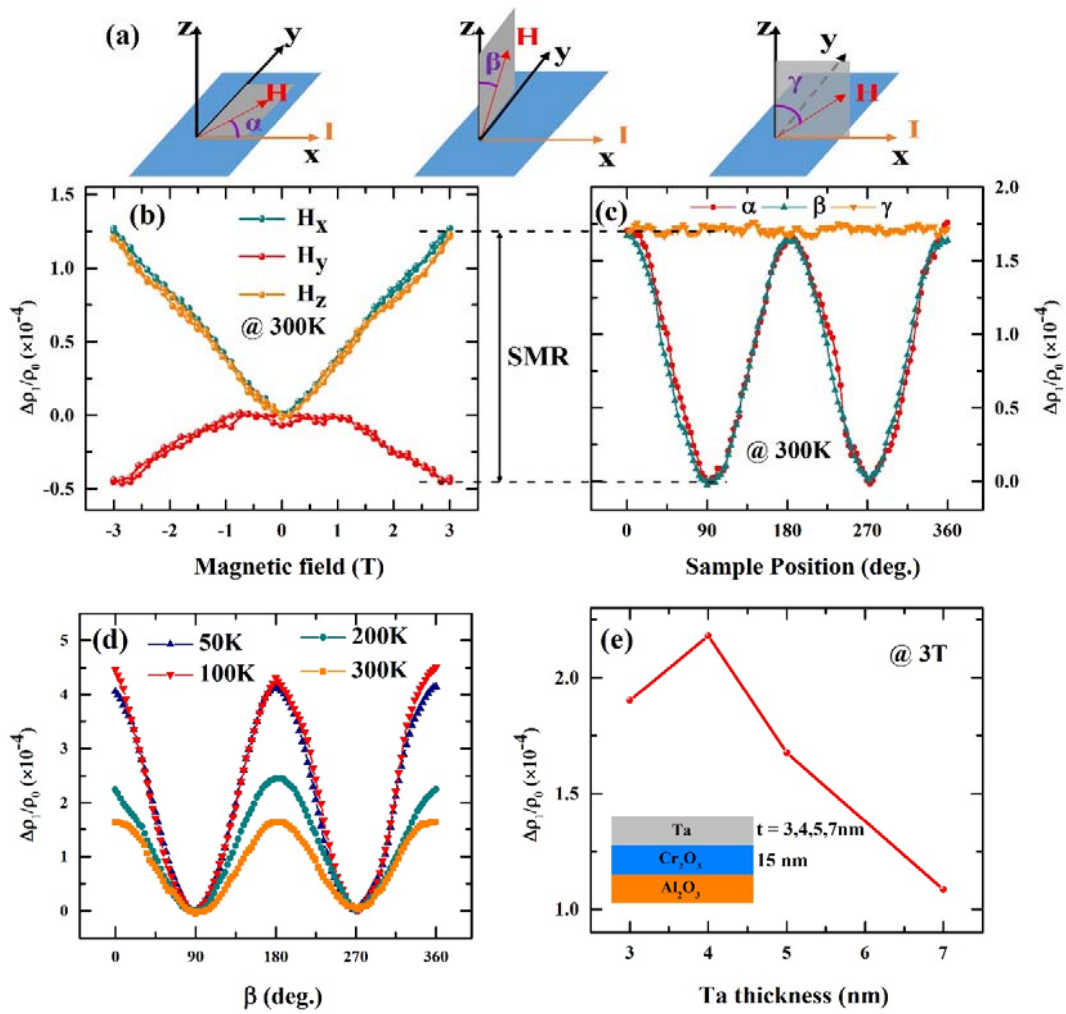

*Figure 2*

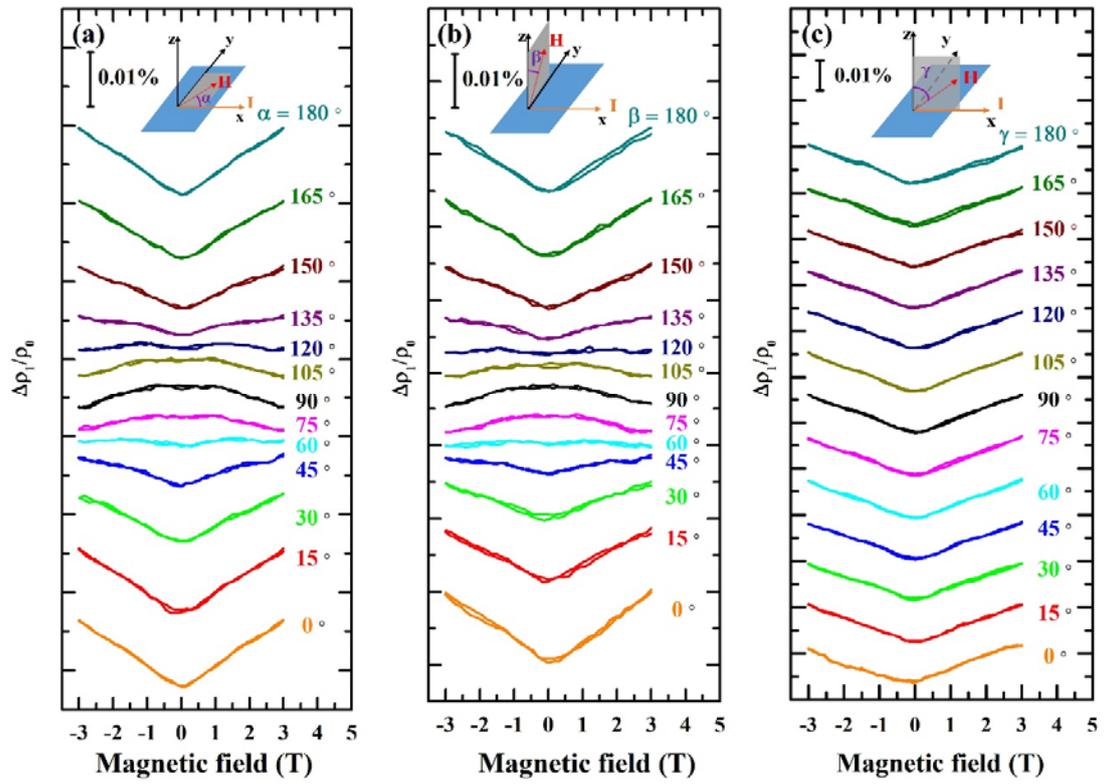

*Figure 3*

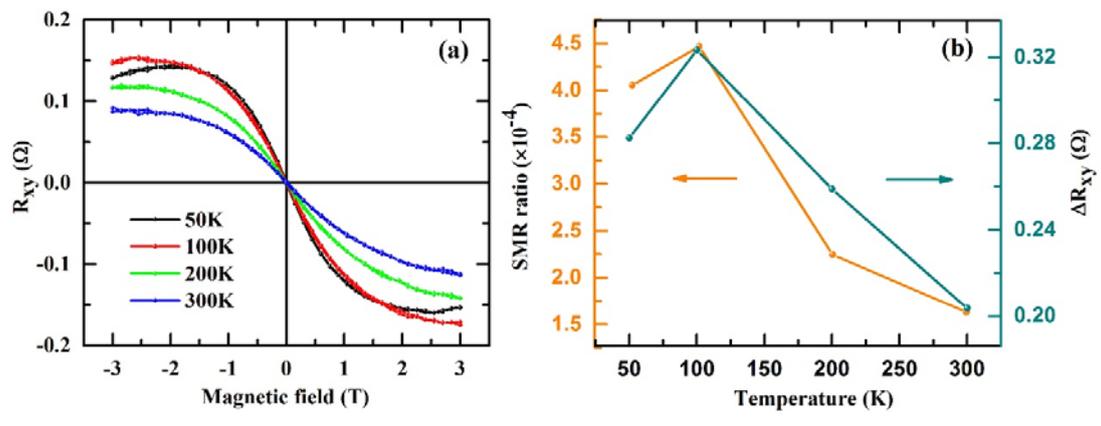

*Figure 4*

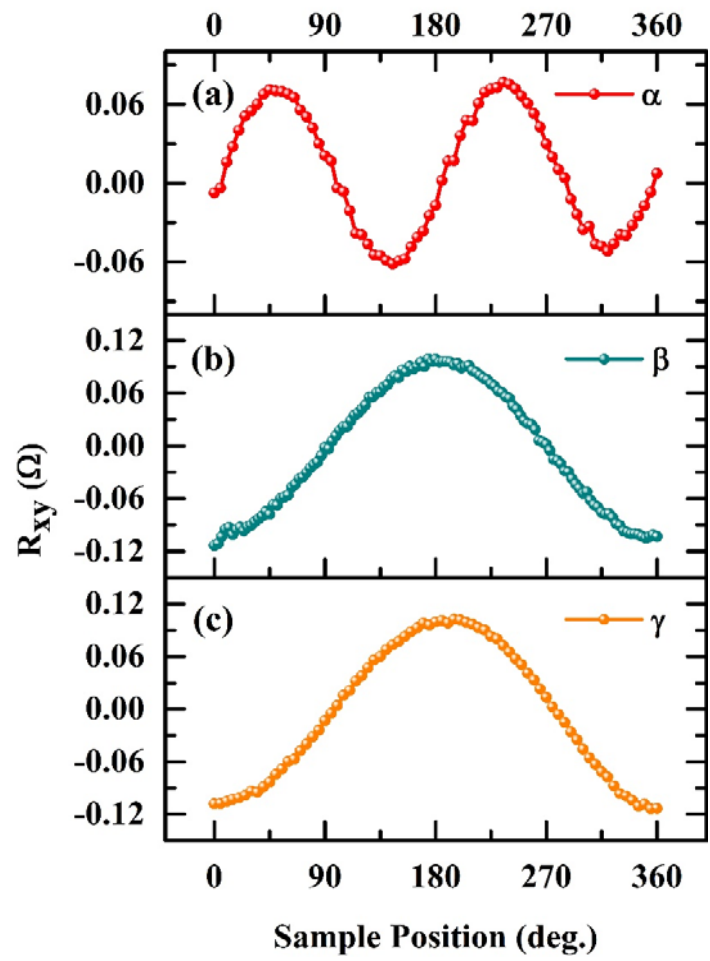

*Figure 5*

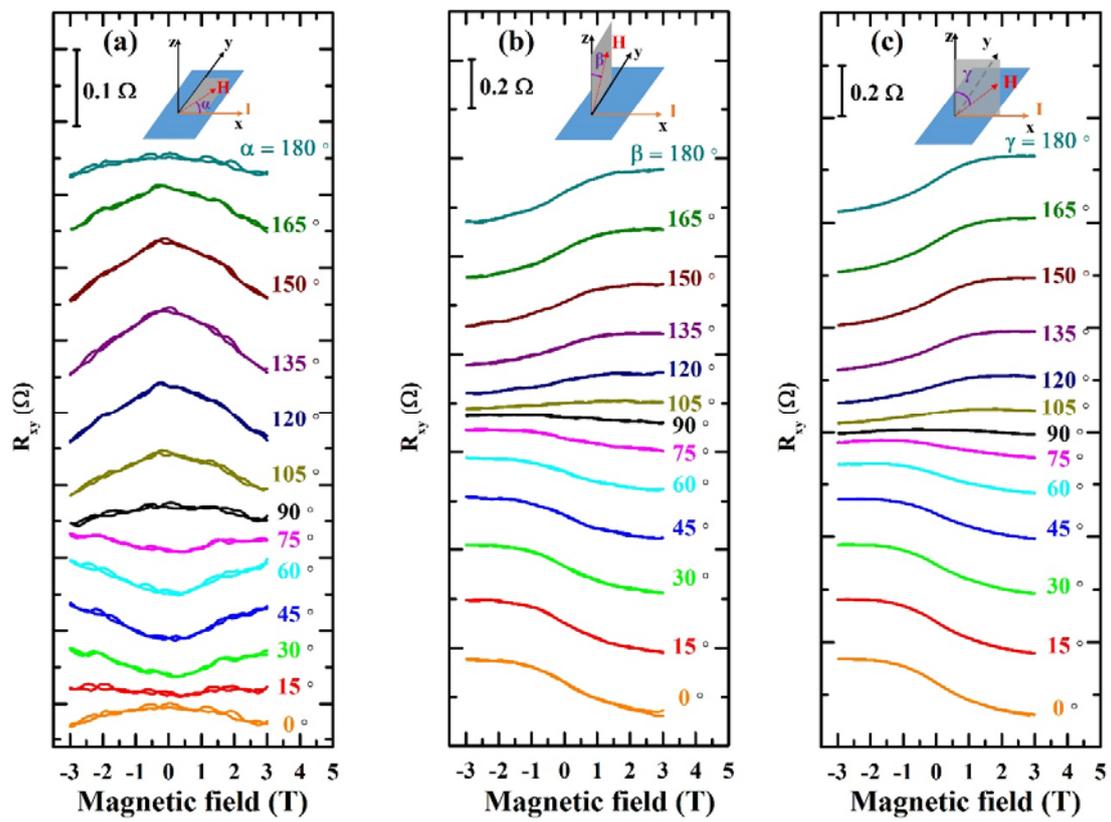

*Figure 6*